\documentclass[11pt,a4paper]{article}
\pdfoutput=1
\usepackage{jcappub}
\usepackage{latexsym,amsmath,amssymb}
\usepackage{slashed}
\usepackage{bm}
\usepackage{mathrsfs}

\newcommand{\nn}{\nonumber}
\newcommand{\bea}{\begin{eqnarray}}
\newcommand{\ea}{\end{eqnarray}}
\newcommand{\rar}{\rightarrow}
\newcommand{\Rar}{\Rightarrow}

\def\dd{\text{d}}
\def\d{\partial}
\def\+{\dagger}

\newcommand{\cth}{\vartheta}
\newcommand{\cep}{\varepsilon}

\newcommand{\cA}{{\cal A}}
\newcommand{\cB}{{\cal B}}

\newcommand{\cH}{{\cal H}}
\newcommand{\cL}{{\cal L}}
\newcommand{\cP}{{\cal P}}
\newcommand{\cQ}{{\cal Q}}
\newcommand{\cS}{{\cal S}}

\newcommand{\vk}{{\mathbf k}}
\newcommand{\vp}{{\mathbf p}}
\newcommand{\vq}{{\mathbf q}}

\newcommand{\vA}{{\mathbf A}}

\newcommand{\vx}{{\mathbf x}}

\def\la{\langle}
\def\ra{\rangle}

\title{The anisotropy of a three- and a one-form}

\author{Federico~R.~Urban}

\affiliation{Service de Physique Th\'eorique, Universit\'e ́Libre de Bruxelles, CP225, Boulevard du Triomphe, B-1050 Brussels, Belgium}

\date{\today}

\abstract{We calculate the anisotropic signal associated with the coupling of a three-form with an Abelian vector gauge field.  In the simplest examples of three-form inflation the amplification of the vector fluctuations is exponential; this makes it almost certain that a large anisotropy will develop, severely constraining the viability of the coupling.}

\arxivnumber{}

\begin{document}
\maketitle

\section{Preamble \& Result}\label{s:intro}

In the Standard model of Cosmology, the seeds for the density perturbations responsible for the inhomogeneous Universe we inhabit are generated during inflation (see for instance~\cite{Mukhanov:2005sc}).  Typically, inflation is a very isotropic affair, owing to the scalar nature of the background field driving the expansion.  There is of course room for interesting extensions to the baseline model, some which include the possibility of generating statistical anisotropies in the final curvature perturbations spectrum.  Interesting in its own right nonetheless, after the results of Planck~\cite{Ade:2013nlj}, hinting indeed at a statistically anisotropic spectrum, such possibility garnered extra credit.

In this work we calculate the statistical anisotropy generated by coupling a three-form to a $U(1)$ gauge field.

Three-fields are among the competitors for ``the'' model of inflation~\cite{Koivisto:2009ew,Germani:2009gg}; they support a perfectly isotropic background~\cite{Koivisto:2009sd,DeFelice:2012jt}; they inflate easily and exhibit some very peculiar dynamics~\cite{Koivisto:2009fb}; they can drive a late-time ``dark energy'' era with peculiar matter interactions~\cite{Ngampitipan:2011se,Koivisto:2012xm}; and, last but not least, they support couplings with a variety of other forms, including the obvious one-form case.

The joint presence of three- and one-forms had been shown to be especially suitable for magnetogenesis models~\cite{Koivisto:2011rm}, for it encompasses the possibility of selecting a narrow range of modes which are amplified by the background expansion, leaving all others untouched in their vacuum state.  Thence, the final power can be distributed entirely among the modes of interest (usually the largest scales), preventing the overall energy density from backreacting onto the de Sitter expansion --- an old plague of inflationary magnetogenesis attempts~\cite{Demozzi:2009fu,Kanno:2009ei,Urban:2011bu}.

A vector field which is dynamic during inflation can source statistically anisotropic classical curvature perturbations~\cite{Bartolo:2012sd,Lyth:2013sha}.  At each horizon crossing, the given curvature perturbation mode will encounter a (subdominant) classical background of the infrared vectors which had left before it.  This background is defined by a preferred direction for each realisation of the inflationary mechanism; direction which is drawn from a given distribution, typically Gaussian.  This was firstly computed for the $I^2(\eta) F^2$ model, with $F_{\mu\nu}$ the $U(1)$ field strength, and $I(\eta)$ its running coupling.

The curvature perturbation evolves in this slightly anisotropic background and its final two-point function (power spectrum) will be explicitly corrected with directional dependence:
\[
\la \zeta_\vp(\eta) \zeta_\vq(\eta) \ra = 2\pi^2 \cP(p) \left[ 1 + \delta\cP(\vp) \right] \frac{\delta^3(\vp+\vq)}{p^3} \, .
\]
We compute this two-point function for a simple model of three- plus one-form where nearly everything can be done analytically~\cite{Urban:2012ib}.  Albeit similar in scope, the structure of the three-to-one-form coupling is very different from that of more common direct inflaton-vector ones, as we will see below.

Before digging the details of the calculation, which can be quite tedious and lengthy, we collect our main findings here.

The gauge field fluctuations are boosted exponentially by the indirect coupling with the background.  This extreme efficiency however is a curse from the point of view of the perturbations, as the anisotropic correction which arises is rapidly enhanced.  We obtain, for the anisotropic (label $\bar{A}$) and isotropic (label $\cA$) contributions:
\bea
\la \zeta_\vp \zeta_\vq \ra^\text{end}_{\bar{A}} &\sim& 8\pi^5 \left( \frac{\rho}{\epsilon\bar{\rho}} \right)^2 \frac{\cH^8}{\Gamma^4 p^4 p_\Lambda^4} \sin^2\cth \, \frac{\delta^3(\vp+\vq)}{p^3} \label{zetabA} \, , \\
\la \zeta_\vp \zeta_\vq \ra^\text{end}_\cA &\sim& \frac{\pi^4}{8} \left( \frac{\rho}{\epsilon\bar{\rho}} \right)^2 \frac{\cH^8}{p_\Lambda^8} \left( \frac13 + \frac{\Gamma p}{p_\Lambda} + \ldots \right) \frac{\delta^3(\vp+\vq)}{p^3} \label{zetacA} \, ,
\ea
where for $\rho$ and $\bar{\rho}$ we mean the energy densities associated with the gauge field and the background three-form, respectively; $\epsilon$ is the first slow-roll parameter $\cH^2-\cH' \equiv \epsilon\cH^2$, where $\cH \equiv a'/a$ is the comoving Hubble parameter, $a(\eta)$ is the scale factor, and priming stands for conformal time $\eta$ derivative.  The $\Gamma$ controls the speed of the enhancement for the vector $U(1)$ field; $p_\Lambda$ is the UV cutoff beyond which modes are not processed by inflation.  Finally, $\cth$ is the angle between $\vp$ and the direction picked by the vector field $\bar{A}$.

Two features need to be discussed.  First, the spectral dependence is such that the anisotropic contribution is greatly enhanced at low momenta.  This is analogous to the scale-invariant case albeit with a very different structure --- the enhancement in that case is only logarithmic.  The reason behind this is that the ealier modes spend more time outside the horizon, where all the action takes place, and will inherit most of the anisotropy.

Second, the overall amplitude.  This is to be compared to the typical curvature perturbation generated by the inflaton itself, which is of order $10^{-9}$.  The gauge contribution is much beyond what is expected for a small correction, and certainly very hard to reconcile with the limits / potential detection from Planck.  Even for very subdominant energy density, the enhancement which comes from the spectral shape $1/p^4$ is too large for the large scale modes.

Notice that in principle the isotropic contribution can be made safe, although this implies that the UV cutoff is at least of order $\cH$ at the end of inflation --- not the ideal situation for magnetogenesis, as we will see.  However, physically it is the anisotropic part which matters: there is no measurable isotropic contribution as computed in~(\ref{zetacA}), as this refers to an infinite number of realisations of the mechanism, and not just the one which we observe.  For a 60 e-folds inflation, the energy density of the $U(1)$ field would have to be suppressed by some $10^{19}$ orders of magnitude.

These result apply to a specific class of models which we describe below.  The main motivations for choosing these models are mathematical simplicity (nearly everything can be done analytically) and the fact that their dynamics are expected to adhere the most to what is needed from inflation (long de Sitter expansion, slow-roll).  However, these are not the only possibilities; one further direction comes from all the additional coupling terms a non-gauge one-form can support.  Also, dynamics further away from fixed points can be envisaged.  The resulting evolution would change quite significantly the picture sketched here.

To summarise, the three-form and the gauged one-form, when coupled, can give rise to highly pronounced anisotropic power spectra, even for strongly subdominant energy density ratios.  While this would ban possible applications to magnetogenesis, it is a very peculiar characteristic of this type of models, and the prominent features of the curvature power spectrum can be easily told apart from other models.

\section{Three- and one-forms}\label{s:onethree}

Our starting point is the action of the system.  Let $A^\mu$ be the $U(1)$ vector potential and $B^{\mu\nu\rho}$ the three-form.  The canonical Lagrangian including both fields is
\bea\label{bare}
\cL_A+\cL_B = -\frac14 F^2(A)-\frac{1}{48} F^2(B) - V(B^2) \, ,
\ea
where the Faraday forms are computed from an $n$-form potential $N$ as $F(N)_{\mu_1\dots\mu_{n+1}}=(n+1)!\d_{[\mu_1}N_{\mu_2\dots\mu_{n+1}]}$.  The components of the dual of the three-form are
\bea
B_\alpha \equiv \frac{1}{6}\epsilon_{\alpha\beta\gamma\delta} B^{\beta\gamma\delta} \, .
\ea
The fully antisymmetric rank 4 tensor is defined by $\epsilon_{0123} = \sqrt{-g}$, $g_{\mu\nu}$ being the metric.  The only minimal coupling term which can be written down which respects the Abelian symmetry of the one-form field reads, for constant $\lambda$,
\bea\label{inte}
\cL_{AB} &=& -\frac{1}{2}\lambda F_{\mu\nu}(A)F^{\mu\nu}(B) \, .
\ea

Working in the Lorenz-Coulomb gauge for which $\d_i A^i = 0 = A_0$, we expand the spatial part of the vector potential in terms of its transverse and longitudinal components as $A_i=A^T_i+\d_i A^L$, where $\nabla\cdot {\vA}^T = 0$.  The three-form is similarly decomposed:
\bea\label{threedec}
B_{0ij} = a^3 \epsilon_{ijx} \left( B_T^k + \d^k B_L \right) \, , \quad B_{ijk} = a^3 \epsilon_{ijk} \left( X - B_0 \right) \, ,
\ea
where the background field is identified with $X=\sqrt{B_\alpha B^\alpha}$, and drives the de Sitter expansion.  The most important structural difference of this model compared to usual direct couplings of the inflaton, is that the $X$ field does not appear in the $F(A) F(B)$ term: three-form and one-form meet through their perturbations, which in turn are coupled to vector metric perturbations, whose dynamics is governed by the background field (see the Appendix).

The only dynamical degrees of freedom are the tranverse ones.  It is possible to obtain a closed equation for the three-form perturbation in Fourier space for momentum $k$ (omitting the transverse tag from now on):
\bea\label{eomB}
\cB_k'' + \left( 1 + \frac{1}{f k^2} \right) k^2 \cB_k = 0 \, ,
\ea
where
\bea\label{fdef}
f = 2\lambda^2 \frac{X}{V_{,X}} \left( \frac{2V_{,X}X}{k^2} - 1 \right) \, ,
\ea
and $\cB_k$ is the Fourier transform of $B_i$; similarly we'll define $\cA_k$:
\bea
A_i(\vx,\eta) \equiv \bar{A}_i(\eta) + \sum_\alpha \int \frac{\dd^3k}{(2\pi)^3} \cep^\alpha_i(\vk) e^{i\vk\cdot\vx} \left( a_\alpha(\vk) \cA_k(\eta) + a_\alpha^\+(-\vk) \cA_k^*(\eta) \right) \, , \label{cAf} \\
B_i(\vx,\eta) \equiv \bar{B}_i(\eta) + \sum_\alpha \int \frac{\dd^3k}{(2\pi)^3} \cep^\alpha_i(\vk) e^{i\vk\cdot\vx} \left( b_\alpha(\vk) \cB_k(\eta) + b_\alpha^\+(-\vk) \cB_k^*(\eta) \right) \, . \label{cBf}
\ea
The $\cep^\alpha_i$ are the polarisation vectors.  We account for the possibility that a homogeneous background for $A_i$ and $B_i$ develops.  This will actually happen due to the mechanism outlined above.

In general the quantity $f$ is time-dependent.  However, if we want a long-lasting inflationary epoch, the background three-form scalar $X$ has to be sitting relatively close to one of its fixed points, where its dynamics are controlled by the potential, and where it will sit for a while.  Thence, we consider $f_0 \equiv f(X_\text{fp})$, and define $\Gamma^2+1 \equiv 1/(f_0k^2)$.  An independent equation for the gauge vector field is also attainable for the gauge field, and reads
\bea\label{eomA}
\cA_k'''' - \left( \Gamma^2 -1 \right) k^2 \cA_k'' - \Gamma^2 k^4 \cA_k = 0 \, .
\ea

The solutions are
\bea
\cA_k(\eta) &=& C^\cA_1 e^{\Gamma k\eta} + C^\cA_2 e^{-\Gamma k\eta} + C^\cA_3\cos{(k\eta)} + C^\cA_4\sin{(k\eta)} \, , \label{cAsolf} \\
\cB_k(\eta) &=& C^\cB_1 e^{i\tilde{\Gamma} k\eta} + C^\cB_2 e^{-i\tilde{\Gamma} k\eta} \, , \label{cBsolf}
\ea
where the $C^\cA_{1,\ldots,4}$ and $C^\cB_{1,2}$ are constants and $\tilde{\Gamma} \equiv \sqrt{\Gamma^2+2}$.  The initial conditions for the system are as usual: Bunch-Davies vacuum in the infinitely remote past $k\eta \rar -\infty$.  Notice that in principle for either $\cA$ or $\cB$, depending on the sign of $\Gamma^2$, it would not be possible to set appropriate initial conditions.  However, in recalling that we are working with a fixed points for $X$, we understand that one simply needs to provide a potential for which the $fk^2 \rar \pm\infty \Rar \Gamma^2 \rar -1$ in the remote past.  If this is the case, both fields will be in vacuum deep inside the horizon.

The interesting solutions are thus
\bea
\cA_k(\eta) &=& \frac{1}{\sqrt{2k}} e^{\Gamma (k\eta+1)} \, , \label{cAsol} \\
\cB_k(\eta) &=& \frac{2}{\sqrt{2k}} e^{-i\tilde\Gamma (k\eta+1)} \, , \label{cBsol}
\ea
where the match with the vacuum solutions is performed at horizon exit.

To be concrete at this point we can specify an archetypical model to work with: the exponential potential $V=V_0\exp{\left( - \xi X^2 \right)}$.  There are two fixed points, $\text{fix}_1$ where $X=0$, and $\text{fix}_2$ where $X=\pm\sqrt{2/3}$ --- the reason why this is a critical point will be clear shortly.   At the first fixed point, which is stable as long as $\xi$ is positive, we have
\bea\label{gamma}
\Gamma^2 = \frac{k_\Lambda^2 - k^2}{\Lambda^2 k_\Lambda^2 + k^2} \, ,
\ea
where we have defined $\Lambda^2 \equiv 8\lambda^2/3 / (1 - 8\lambda^2/3) \simeq 8\lambda^2/3$ and $k_\Lambda^2 \equiv 8\xi V / (3\Lambda^2) \simeq \xi V / \lambda^2$.  Only for $k \leq k_\Lambda$ there exist an exponentially growing solution for $\cA$: $k_\Lambda$ acts in this case at the \emph{de facto} UV cutoff of the $U(1)$ theory.

The second fixed point can be an attractor if $\xi$ is negative.  In this case one obtains
\bea\label{gammanot}
\Gamma^2 = \frac{\xi V_0 - \lambda^2 k^2}{\lambda^2 k^2} \, .
\ea
But here, when this point is an attractor, there is no instability for the $U(1)$ field.  However, at this point, it is $\cB$ which starts growing.  If instead we are inflating at a saddle point (i.e., $\xi>0$), then gauge vector modes will be unstable at scales $\lambda^2k^2<\xi V_0$, and three-form rotational modes will remain in the vacuum.  An interesting phenomenon is in action here: the three- and one-forms are \emph{mutually exclusively} amplified.  The non-trivial dynamics spring from both the existence of a coupling, and the three-form having a potential.   Thanks to the duality between the dual three-form field and the gauge field, it would be in general possible to give a potential to the one-form to drive the system, but this would make it inherently anisotropic.  We leave this possibility for future work.

Since the dynamics is very similar for the two vector fields, we focus on the more interesting case of the gauge field; it should be borne in mind however that the results we obtain could be translated directly to the three-form rotational modes directly, for appropriate choice of the potential.

The energy density stored by the $U(1)$ field can be computed as:
\bea\label{energy0}
\rho_\cA = \frac{1}{4\pi^2 a^4} \int \frac{\dd k}{k} k^3 \left[ |\cA'|^2 + k^2 |\cA|^2 \right] \, .
\ea
In order to be able to work within perturbation theory we keep $\lambda$ small, which means $\Lambda\lesssim1$ and $\Gamma\gtrsim1$; if we take $\Gamma = \text{const}$ up to some given cutoff $k_\text{cut}$ for simplicity then
\bea\label{energy}
\rho_\cA \simeq \frac{k_\text{cut}^4}{32\pi^2 a^4} \Gamma^2 e^{2\Gamma} \, .
\ea
Notice how the cutoff scale $k_\text{cut}$ appears in this expression: high energy modes are left untouched by the de Sitter expansion.  The UV cutoff can be approximated as $k_\Lambda / \Gamma$, since due to the shape of $\Gamma$, the integrand starts falling off with momentum earlier than the actual cutoff at $k_\text{cut}$; within the order of magnitude results we are attaining, this is a fair approximation.  We employ both results later to estimate the two-point curvature perturbation.

\section{The curvature perturbation}\label{s:curvature}

We want to understand the effect of the dynamical evolution of the gauge field on the gravitational scalar curvature perturbation $\zeta$.  The latter is defined as a combination of the metric perturbations in a given gauge.  For instance, in longitudinal gauge
\bea\label{metric}
\dd s^2 = a^2 \left[ (1+2\Phi) \dd \eta^2 - (1-2\Psi) \dd \vx^2 \right] \, .
\ea
We derive the perturbed Einstein field equations $\delta G_{\mu\nu} = \delta T_{\mu\nu}$ (in units where the gravitational coupling constant is 1) again in Fourier space as
\bea
&& 3\cH \Psi' + 3\cH^2 \Phi + k^2 \Psi = -\frac12 a^2 \left( \delta\bar{\rho} + \rho \right) \, , \label{psip}\\
&& \Psi''+ 2\cH \Psi' + \cH \Phi' + (2 \cH' + \cH^2) \Phi - \frac{k^2}{3} (\Phi-\Psi) = \frac12 a^2 \left( \delta\bar{P} + P \right) \, , \label{psidp}\\
&& \Psi' + \cH \Phi = - \frac12 \frac{a \, {\rm i} \, k^j}{k^2}\, \left( \delta\bar{q}_j + q_j \right) \, , \label{psiq}\\
&& k^2 (\Phi-\Psi) = -a^2 \Pi \, , \label{phimpsi}
\ea
where $\Pi(\vk)\equiv -3 \hat k^i\hat k_j \Pi^j_i(\vk)/2$, and where all perturbed quantities come from the stress-energy tensor, decomposed with respect to the fluid velocity $u^\mu =(1-\Phi,{\bf 0})/a$:
\bea\label{Tmunu}
T^{\mu\nu}=(\rho +P) u^\mu u^\nu - P g^{\mu\nu} + u^\mu q^\nu + u^\nu q^\mu + \Pi^{\mu\nu} \, .
\ea
The barred set $(\bar{\rho},\bar{P},\bar{q},\bar{\Pi})$ refers to the three-form (background and, with a $\delta$ attached, perturbations) while the unbarred ones belong to the gauge field.  Both act as sources for the metric perturbations, and in turn, to the curvature perturbations $\zeta$.  Notice that we are disregarding second order perturbations in the metric and the three-form, for they are subdominant in the present setup.  Finally, the background equations which govern the evolution of $X$ are
\bea
&& X'' + 2\cH X' + 3\cH' X - 3\cH^2 X + a^2 V_{,X} = 0 \, ,\label{xeom}\\
&& \cH^2 = \frac13 \left( \frac12 K'^2 + a^2 V_{,X} \right) \, ,\label{fx1}\\
&& \cH^2 - \cH' = \frac12 a^2 V_{,X} X \, .\label{fx2}
\ea
Here $K' \equiv X' + 3\cH X$.  It is from these equations that we see why there is a new fixed point, pertaining to the three-form only, corresponding to $X^2=\pm2/3$: at this value the background Friedmann equation saturates, and no further development takes place.

The curvature perturbation $\zeta(\vk,\eta)$ in this gauge is
\bea\label{zetadef}
\zeta \equiv \Psi + \frac{\cH^2}{\cH^2-\cH'} \left( \frac{\Psi'}{\cH} + \Phi \right) \, .
\ea
We find it convenient to work in this gauge because it is relatively easy to obtain a closed equation for the exact curvature perturbation $\zeta$, see below.  One can certainly work in more common gauges (flat gauge, uniform density gauge), and obtain the same result.  The dynamics of $\zeta$ are dictated by
\bea\label{zetap}
\zeta' = -\frac{\cH}{\epsilon} \left\{ \left( \frac{c_s k}{\cH} \right)^2 \Psi + \frac{1}{\bar{\rho}} \left[ \Pi + \frac32 \left( c_s^2 \rho - P \right) - 3 \left( \frac32(c_s^2-1) - \epsilon + \frac{\epsilon'}{2\cH\epsilon} \right) Q \right] \right\} \, ,
\ea
with
\bea\label{Qdef}
Q \equiv {\rm i}\frac{\cH k^j q_j}{k^2 a} \, .
\ea
The flow $Q$ is related to the other quantities in the source via momentum conservation:
\bea\label{Qcons}
\frac{Q}{\cH} = \frac23 \Pi - P - (3+\epsilon) Q \, ;
\ea
we will also use the fact that, of course, $P = \rho/3$ for a massless field.  All these expressions are exact --- we used the slow roll parameter $\epsilon \cH^2 = \cH^2 - \cH'$ only for notational convenience.  Then, within the slow-roll approximation $\epsilon\ll1$ (the three-form also inflates by slowly rolling along its potential, although oftentimes climbing it rather than sliding down), by deriving equation~(\ref{zetap}) with respect to conformal time, and using the Einstein and momentum conservation equations, we finally obtain
\bea\label{zetadyn}
\zeta'' + 2\cH \zeta' + c_s^2 k^2 \zeta = -\frac{\cH}{\epsilon\bar{\rho}} \left\{ \Pi' + \frac12 (3c_s^2-1) \rho' + 6c_s^2 \cH \rho \right\} \, ,
\ea
with the speed of sound $c_s^2 \equiv V_{,XX} X / V_{,X}$ --- in the specific case of a type 1 fixed point for the exponential potential the latter is just 1.

Eq.~(\ref{zetadyn}) can be solved using the Green's function method (equivalent to the Wronskian method).  It is convenient to rewrite it in terms of\footnote{Note: not to be confused with the configuration space coordinate $\vx$.} $x\equiv-k\eta$ and using $\cH = -1/\eta$:
\bea\label{zetax}
\zeta_{,xx}(x) - \frac{2}{x} \zeta_{,x}(x) + c_s^2 \zeta(x) = \cS(x) \, ,
\ea
where the source term is
\bea\label{source}
\cS(x) = \frac{1}{\epsilon\bar{\rho}} \left[ x\Pi_{,x} + x\rho_{,x} - 6c_s^2\rho \right] \, .
\ea
Integrating over ``time'' $x$:
\bea
\zeta_\vk(x) &=& \int_{x_h}^x \dd y G(x,y) \cS_\vk(y) \, , \label{solzeta} \\
G(x,y) &\equiv& \frac{\pi}{2} y \left(\frac{x}{y}\right)^{3/2} \! \left[ J_{3/2}(y)J_{-3/2}(x)-J_{3/2}(x)J_{-3/2}(y) \right] \, . \label{green}
\ea
The $J_n(x)$ are Bessel functions, and the entire Green's function can be simplified for small $(x,y) \ll 1 \approx x_h$, as we work with superhorizon modes\footnote{This assumption needs to be checked \emph{a posteriori}, that is, the most important contribution to the integral must come from the final time $\eta_\text{end}$, for which $k|\eta_\text{end}|\ll1$.}, to $-y/3$.

What we need to compute is therefore the two-point function
\bea\label{twop}
\la \zeta_\vp(x) \zeta_\vq(x) \ra &=& \int \dd y \dd z \, G(x,y) G(x,z) \la \cS_\vp(y) \cS_\vq(z) \ra \nn \\
&\approx& \frac19 \int \frac{\dd y}{y} \frac{\dd z}{z} \la y^2 \cS_\vp(y) z^2 \cS_\vq(z) \ra \, .
\ea

The source term, given our exponential solutions for the gauge potentials, and given that we expect $\Gamma\gtrsim1$, is going to be dominated by the conformal time derivative terms $\cA'(x)$ and $\bar{A}_i'(x)$; we thus retain only these terms in $\Pi$ and $\rho$ which become
\bea
\Pi_\vk^{\bar{A}}(x) &\subset& -\frac{1}{a^4} \bar{A}_i'(x) \sum_\alpha \varepsilon_i^\alpha(\vk) a_\alpha(\vk) \cA_k'(x) + \text{h.c.} \, , \label{pibA} \\
\Pi_\vk^\cA(x) &\subset& -\frac{1}{2a^4} \sum_{\alpha\alpha'} \int\frac{\dd^3 k'}{(2\pi)^3} \cep^\alpha_i(\vk') \cep^{\alpha'}_i(\vk-\vk') a^\+_\alpha(\vk') a^\+_{\alpha'}(\vk-\vk') \cA_{k'}'(x) \cA_{|\vk-\vk'|}'(x) + 3\text{h.c.} \, , \label{picA} \, .
\ea
Next is the two-point function of the anisotropic stress (and the energy density $\rho\sim3\Pi$), which we can define, splitting it into a time-dependent part and the integrals over momenta, as
\bea
\la \Pi_\vp(y) \Pi_\vq(z) \ra_{\bar{A}} &=& \frac{(2\pi^3)\delta^3(\vp+\vq)}{a_y^4a_z^4} \left[\bar{A}_i'(y)\bar{A}_i'(z) - \hat{p}\bar{A}'(y) \, \hat{q}\bar{A}'(z)\right] \cA_p'(y) \cA_q'(z) \nn\\
&\equiv& \la\hat{\Pi}_{\bar{A}}^2\ra \delta^3(\vp+\vq) y^4 z^4 e^{-2\Gamma(y+z)} \, , \label{twopbA} \\
\nn\\
\la \Pi_\vp(y) \Pi_\vq(z) \ra_\cA &=& \frac{\delta^3(\vp+\vq)}{2a_y^4a_z^4} \int\dd^3k\left[1+(1+\hat{p}\hat{k})^2\right]^2 \left[\cA_k'\cA_{|\vp+\vk|}'\right]_y \left[\cA_k'\cA_{|\vq+\vk|}'\right]_z \nn\\
&\equiv& \la\hat{\Pi}_\cA^2\ra \delta^3(\vp+\vq) y^4 e^{-2\Gamma \cP y} z^4 e^{-2\Gamma \cQ z} \, , \label{twopcA}
\ea
where $\cP \equiv (k+|\vp+\vk|)/p$ and similarly $\cQ \equiv (k+|\vq+\vk|)/q$ (although in the end $\vp=-\vq$ we keep them explicit for now), and we have used our solutions~(\ref{cAsol}).  The hatted $\hat{\Pi}$ do not depend on time; notice that in the case of $\bar{A}_i(x)$ we have used $\bar{A}_i(x) = \bar{A}_i(x_e) \exp(-\Gamma (x-x_e)) \sim \bar{A}_{e\,i} \exp(-\Gamma x)$, as the background of infrared modes keeps growing as inflation proceeds.  Also, we write $a(x) = k/Hx$ with $aH=\cH$ defining the physical Hubble parameter $H$.

In order to explicitate the source term, we take the coincidence limits of these two-point functions for $\Pi$ as the r.m.s.\ of the fluctuations associated to this operator.  Then, its square root gives the time dependence of $\Pi$ as well as of $\rho$, as $x^4 \exp(-2 \Gamma x)$, modulo some order one $\cP$ and $\cQ$ factors.  This means that, for example, $x\Pi_{,x} \simeq 4\Pi -2\Gamma x \Pi \rar 4\Pi$, for even if $\Gamma\gtrsim1$ we expect $x\ll1$ especially in view of the presence of the UV cutoff $k_\Lambda \ll \cH$ at the end of inflation.

The time integrals can then be performed easily in the limit of large wavelengths, which for instance allows us to replace the integration endpoints as $(y=[x_h,x],z=[y_h,y]) \rar (y=[x_h,x],z=[x_h,x])$ so that the double integral becomes an integral squared --- we then discard the initial contribution as it is small.  Thence:
\bea\label{timeint}
\int\frac{\dd y}{y}\frac{\dd z}{z} y^4 z^4 e^{-2\Gamma(\cP y + \cP z)} &\approx& \frac{1}{2^9} \left(\frac{1}{\Gamma \cP}\right)^8 e^{-2\Gamma\cP x} \left(6 + 6(\Gamma\cP x) + 3(\Gamma\cP x)^2 + (\Gamma\cP x)^3\right)^2 \nn\\
&\approx& \frac{9}{128} \left(\frac{1}{\Gamma \cP}\right)^8 \, .
\ea
The corresponding expression for the contribution to $\la \Pi^2 \ra_{\bar{A}}$ is identical but with $\cP=1$.  Finally, we can approximate the momentum integrals of the $\la \hat{\Pi}^2 \ra$ (not forgetting the $k$-dependence from the time integration) by ignoring the angular dependence, and by taking the $\Gamma$ as constant up to a given UV cutoff $k_\text{cut}$.  This returns the order of magnitude estimates
\bea
\la\hat{\Pi}_{\bar{A}}^2\ra &\approx& \frac{(2\pi)^3 H^8}{2p^8} \Gamma^2 p e^{2\Gamma} \left[\bar{A}_{e\,i}'\bar{A}_{e\,i}' - (\hat{p}\bar{A}_e')^2\right] \, , \label{hatpbA} \\
\la\hat{\Pi}_\cA^2\ra &\approx& \frac{\pi H^8}{p^3} \Gamma^4 e^{4\Gamma} \left(1+3\frac{p}{p_\text{cut}}+\ldots\right)\, . \label{hatpcA} \\
\ea

All in all, the two-point functions now read
\bea
\la \zeta_\vp \zeta_\vq \ra^\text{end}_\cA &\sim& \frac{\pi}{3\times2^{13}} \left( \frac{1}{\epsilon\bar{\rho}} \right)^2 \frac{\cH^8}{\Gamma^4} e^{4\Gamma} \left( 1 + 3\frac{p}{p_\text{cut}} + \ldots \right) \frac{\delta^3(\vp+\vq)}{p^3} \label{zetacApre} \, , \\
\la \zeta_\vp \zeta_\vq \ra^\text{end}_{\bar{A}} &\sim& \frac{\pi^3}{4} \left( \frac{1}{\epsilon\bar{\rho}} \right)^2 \frac{\cH^8}{\Gamma^6 p^4} e^{2\Gamma} \bar{A}_{e\,i}'\bar{A}_{e\,i}' \sin^2\cth \, \frac{\delta^3(\vp+\vq)}{p^3} \label{zetabApre} \, .
\ea
The results quoted in the introduction are obtained by replacing
\[
\bar{A}_{e\,i}'\bar{A}_{e\,i}' \sim 2a^4\rho \sim \Gamma^2 e^{2\Gamma} p_\text{cut}^4 / 16\pi^2 \, ,
\]
and eventually recognising that the actual cutoff is $p_\text{cut} \sim p_\Lambda/\Gamma$.

\section{Analysis \& Alternatives}\label{s:aa}

The expression~(\ref{zetabA}) shows how the perfect efficiency of the three-form in amplifying one-form perturbations results in a wild enhancement of curvature perturbations, to the point where it is not easy to imagine how it would be reconciled with observational demands.  Typically, for magnetogenesis applications, we need low cutoff scale $k_\Lambda$ and not too small $\Gamma$.  These two requirements ensure that the band which is boosted is narrow, and that it is much boosted.

The same choices, when turning to the anisotropic curvature spectrum, carry nefastus consequences.  For low momentum modes there is a drastic enhancement, and only a very strongly subdominant energy density stored in the $U(1)$ field would be viable.  It is easy to convince oneself that, even in the best case scenario where the cutoff is at (or beyond) the highest momentum $\cH$ at the end of inflation, the low energy modes still contribute too much to curvature perturbation.

This appears to be somewhat in contradiction with what was previously found~\cite{Urban:2012ib}.  In that paper the calculation was performed entirely in terms of the longitudinal modes potentials, and only at the end the curvature perturbation $\zeta$ was extracted.  The two methods are equivalent in all aspects; in~\cite{Urban:2012ib} there was a mistake in performing the momentum integrals, and a cancellation among low powers of $k|\eta|$ terms was found; this lead directly to the result quoted there.

These results are strictly only valid for the case of the exponential potential $V=V_0\exp{\left( - \xi X^2 \right)}$; in fact, they generally can be applied to most toy-models of three-form magnetogenesis which admit exact solutions, for the basic features (low energy cutoff, exponential amplification) persist in all of them, and the outcomes will be the same.

The most obvious workaround for this problem is to choose a potential for which the boost is not exponential.  In this case, even for a relatively large power, the time-integrals would return simple powers of $k|\eta|$, and the contribution to $\langle\zeta^2\rangle$ would be accordingly suppressed.  So, compared to the standard magnetogenesis with $f^2F^2$, we would also take advantange of the structural form of the coupling, which we have seen does not implement any slow-roll enhancement for the anisotropic contribution.  The latter would then be made safe much more easily.

Another option is to throw away the gauge-invariance of the one-form, and allow for more general couplings between the two fields.  Terms such as $\nabla_\mu A^\mu \nabla_\nu B^\nu$ or $m^2 A_\mu B^\mu$, and direct couplings to the Ricci and Riemann tensors would be permitted.  The link with magnetogenesis in that case is lost, but there can arise several new signatures for both two- and three-point curvatures.  We leave this possibility for upcoming work.

Independently on magnetogenesis, the result (\ref{zetabA}) shows some interesting features which would allow discrimination between the statistical anisotropy produced with the present mechanism from the other models.  These features are first of all the strong momentum depence of the anisotropic coefficient.  Secondly, the existence of only a narrow window of modes which contribute to the anisotropy, and in turn to non-Gaussianities.  There would then be a strong signal in the lowest end of the spectrum, but nothing would be seen beyond the cutoff.

The results obtained here can be found with different methods as well.  One can easily check that in the so-called in-in formalism the time-integrals take the exact same form, and would return the same answers.  The only subtlety in that case is that, being the direct $(\zeta A' A')$ term absent, the main contributions to the tree and loop diagram interactions come from slow-roll suppressed couplings (through the metric perturbations).

To conclude, three-form inflationary magnetogenesis does not appear to be simply viable as the background analysis would promise.  The model can still be operative and successful but needs further refinements; refinements which can be analysed by numerical methods, since the exact analytical expressions we have derived strictly apply only to the toy-models examples built around the exact stationary point for the background inflaton.

On the other hand, the coupled one-form needs not be the electromagnetic field.  In this case there are first of all additional coupling terms one can write down, whose effects on the dynamics can differ significantly; secondly, there is more freedom in the parameter space of the model, which is then only constrained by backreaction bounds.

Since the three-form has proven to be a healthy contender for inflation, we believe these alternatives are still worth exploring, despite the negative results presented in this work.  We leave all these possible directions of investigation for future research.

\section*{Acknowledgements}
FU wishes to thank Tomi Koivisto, Mindaugas Kar\v ciauskas, Kei Yamamoto, Chris Byrnes, and Lukas Hollenstein for useful discussion and correspondence.  He is supported by the IISN project No.~4.4502.13 and by Belgian Science Policy under IAP VII/37.

\section*{Appendix}

In this appendix we briefly describe the structure of the three-to-one-form coupling, to point out similarities and differences with the more standard direct couplings of the background inflaton with vector gauge perturbations.

Consider first scalar field inflation $X$ coupled to the $U(1)$ field as
\[
  {\cal L} = \sqrt{-g} \left[ I^2(X) F^2 + (\d X)^2 + V(X) + R \right] \, ,
\]
where the last term is the gravitational action.  The fields $g$, $X$, and $A$ are perturbed as $g \rar 1 + \delta g$, $X \rar 1 + \delta X$, and $A \rar \bar{A} + \delta A$ --- this is a schematic expansion whose purpose is to study the structure of the couplings, so we disregard the homogeneous parts, except for the gauge field since we are interested in the anisotropy it produces.

In the slow-roll approximation we can write
\bea
\sqrt{-g} &\rar& 1 + \delta g + \delta g^2 + \ldots \, , \nn\\
I^2 &\rar& 1 + \lambda \delta X + \lambda^2 \delta X^2 + \ldots \, , \nn\\
(\d X)^2 &\rar& \epsilon g (1 + \delta X + \delta X^2) \, , \nn\\
F^2 &\rar& g^2 (\bar{A}^2 + \bar{A} \delta A + \delta A^2) \, , \nn\\
V &\rar& 1 + \sqrt{\epsilon} \delta X + \epsilon \delta X^2 + \ldots \, , \nn\\
R &\rar& g^2 \rar 1 + \delta g + \delta g^2 \, . \nn
\ea
The $\epsilon$ factors account for the slow-roll expansion, and $\lambda \sim 1/\sqrt{\epsilon}$ is the coupling between $X$ and the photon.  One can expand the action and obtain the quadratic part.  Varying it with respect to, for example, the $\delta g$ metric perturbations (which we want to eliminate from the action) we obtain $\delta g = \bar{A} \delta A + (\sqrt{\epsilon} + \lambda \bar{A}^2) \delta X$ at first order.  The quadratic action then becomes
\[
  \delta A^2 + (\lambda + \sqrt{\epsilon}) \bar{A} \delta A \delta X + (\epsilon + \lambda^2 \bar{A}^2) \delta X^2 \, ,
\]
and the cubic piece is
\[
  \bar{A} \delta A^3 + (\lambda + \sqrt{\epsilon}) \delta A^2 \delta X + (\epsilon + \lambda^2) \bar{A} \delta A \delta X^2 + (\epsilon^{3/2} + \lambda^3 \bar{A}^2) \delta X^3 \, .
\]
Clearly the dominant contribution to the $\delta X$ perturbation comes from the $\lambda \bar{A} \delta A \delta X$ mass insertion.  The first loop diagram is the $\lambda \delta A^2 \delta X$, in accordance to what was found in~\cite{Bartolo:2012sd,Lyth:2013sha}.  These diagrams are slow-roll enhanced through the coupling $\lambda$.  In the end, one converts $\delta X$ to $\zeta$ in a given gauge, and obtains the interactions which appear in the in-in formalism.

Now let us turn to the three-form case.  The action is
\[
  {\cal L} = \sqrt{-g} \left[ F_A^2 + F_B^2 + V(B^2) + \lambda F_A F_B + R \right] \, ,
\]
where we will denote $X$ the background three-form scalar as in the main text.  In the slow-roll approximation we can again write
\bea
\sqrt{-g} &\rar& 1 + \delta g + \delta g^2 + \ldots \, , \nn\\
F_B^2 &\rar& \epsilon g^2 (1 + \delta X + \delta X^2) \, , \nn\\
F_A^2 &\rar& g^2 (\bar{A}^2 + \bar{A} \delta A + \delta A^2) \, , \nn\\
\lambda F_A F_B &\rar& \lambda \sqrt{\epsilon} g^2 (\bar{A} \delta X + \delta A \delta X) \, , \nn\\
V &\rar& 1 + (1+\delta g + \delta g^2 + \ldots) [\sqrt{\epsilon} (\delta X + \delta g) + \epsilon (\delta X + \delta g)^2 + \ldots] \, , \nn\\
R &\rar& g^2 \rar 1 + \delta g + \delta g^2 \, . \nn
\ea
The cross-term $F_A F_B$ does not contain the background $X$.  Notice that here $\lambda$ is a simple coupling, and does not hide any slow-roll parameter.  following the same procedure, we obtain $\delta g = (\bar{A} + \lambda \sqrt{\epsilon}) \delta A + \sqrt{\epsilon}(1 + \lambda \bar{A}) \delta X$ at first order.  The quadratic action then becomes
\[
  \delta A^2 + \sqrt{\epsilon}(\lambda + \bar{A}) \delta A \delta X + \epsilon \delta X^2 \, ,
\]
and the cubic one is
\[
  (\lambda \sqrt{\epsilon} + \bar{A}) \delta A^3 + \sqrt{\epsilon}(\lambda \bar{A} + 1) \delta A^2 \delta X + \epsilon(\bar{A} + \lambda) \delta A \delta X^2 + \sqrt{\epsilon^3}(\lambda \bar{A} + 1) \delta X^3 \, .
\]
The same diagrams which in the scalar field case were dominating now do not come directly from the $F_A F_B$ coupling, but from the indirect coupling to the metric perturbations, and are therefore slow-roll suppressed: $\sqrt{\epsilon} \bar{A} \delta A \delta X$ and $\sqrt{\epsilon} \delta A^2 \delta X$.  The first diagram to contain the coupling $\lambda$ which contributes to the $\la \delta X^2 \ra$ two-point function is the $\sqrt{\epsilon} \bar{A} \delta A^2 \delta X$ term, which has higher powers of the subdominant $\delta A$ field, on top of the remaining slow-roll suppression.  In this case one can easily show that the most important contribution to the \emph{curvature} perturbation $\zeta$ does not come from $\delta X$ directly, but rather from the plain energy density of the $U(1)$.  This is in agreement with what we have found above in the longitudinal gauge.

\bibliography{3fBany}
\bibliographystyle{JHEP}

\end{document}